\documentclass[runningheads]{cl2emult}

\usepackage{makeidx}  
\usepackage{graphicx} 
\usepackage{subeqnar} 
\usepackage{multicol} 
\usepackage{cropmark} 
\usepackage{eso}      
\makeindex            

\begin{document}
\title*{Where are the First Stars now?}
\titlerunning{Where are the First Stars now?}
%
\author{Simon D.M. White \& Volker Springel}
\authorrunning{White \& Springel}
%
%
\institute{Max-Planck-Institute for Astrophysics, Garching bei
M\"unchen, Germany}

\maketitle              

\begin{abstract}
We use high-resolution simulations to show that the current standard
paradigm for the growth of structure in the Universe predicts 
the formation of a galaxy like our own to differ substantially from the
classic ELS and Searle/Zinn pictures. On scales larger than the Local 
Group, the earliest star formation was extremely inhomogeneous, 
suggesting that the high redshift intergalactic medium should have 
large-scale abundance variations. The very oldest stars should be 
found today in the central regions of rich galaxy clusters. In the 
Milky Way's bulge and stellar halo little correlation is expected 
between age and metallicity. Some of the lowest metallicity stars may 
be relatively young. Many of the oldest stars may have high
metallicity. Spheroid stars were formed before and during halo assembly, the
oldest now lying preferentially at small radii, while low metallicity 
stars lie preferentially at large radii. The bulk of the Milky Way's 
stellar spheroid came from a small number of progenitors. It should
show little spatial structure in the inner 5 to 15 kpc, but consist
at each point of a superposition of hundreds of `cold' streams. Such
streams have been detected in the Solar neighborhood.
\end{abstract}

\section{Introduction}

Over the last two decades a standard paradigm has emerged for the
origin and evolution of structure in the Universe. The basic
ingredients of this picture are the following.\hfil\break
$\bullet$  The Universe has expanded from a hot, dense and smooth
state. This Hot Big Bang is observed directly in the CMB at an 
age of 300,000 years. It is checked by the Planckian nature of
the CMB spectrum back to an age of a few months, and by the abundances
of the light elements back to an age of a few minutes.\hfil\break
$\bullet$  The observed near-homogeneity and isotropy were produced by
an early phase of inflationary expansion.\hfil\break
$\bullet$  The dominant matter component today is some unseen form of
nonbaryonic, weakly interacting dark matter.\hfil\break
$\bullet$  All observed structure originated from quantum zero-point
fluctuations during inflation. These produced a gaussian field of
density fluctuations with a near-Harrison-Zel'dovich power spectrum
showing no characteristic features on scales relevant to galaxies and
larger structures.\hfil\break
$\bullet$  Present structure grew by gravitational amplification of
these small initial fluctuations.\hfil\break
$\bullet$  Galaxies formed by cooling and condensation of gas in the 
cores of heavy halos produced by nonlinear hierarchical
clustering of the dark matter.\hfil\break

The current ``consensus'' version of this model imagines a flat
Universe in which baryons contribute a few percent of the closure
density, cold dark matter contributes about one third, and an
effective cosmological constant contributes the rest. This model
is consistent with current data on the distance scale,
acceleration, large-scale structure and galaxy populations of the
Universe. It should be definitively tested by the next
generation of CMB experiments, in particular by the MAP and Planck
Surveyor satellites.

Strong points of such CDM models are that they provide fully specified
initial conditions, and that the nonlinear growth of structure in the
dominant component, the dark matter, can be simulated accurately on 
scales larger than a kiloparsec or so. Below this scale nongravitational 
processes undoubtedly play a major role in shaping galaxies. Recent 
work has grafted phenomenological modelling of gas cooling, 
star-formation, feedback, and stellar evolution onto high resolution 
N-body calculations of dark matter clustering \cite{KCDW}. This has made it possible 
to simulate cosmologically representative volumes with sufficient 
resolution to follow the formation and evolution  of individual galaxies. 
Here we present preliminary results from an extension of this work to 
much higher resolution. We trace the detailed formation history of 
individual objects and of their immediate environment, showing results 
for a rich galaxy cluster and for a system like the Milky Way. A much
more detailed account of the results for clusters will be published
shortly as Springel et al (in preparation).

\section{The Simulations}

The simulations discussed here were carried out using a parallel
tree-code called Gadget \cite{Gadget} on the Cray T3E at the Garching Computing
Centre of the Max Planck Society. Initial conditions were taken from
a $\Lambda$CDM simulation already analysed by Kauffmann and
collaborators \cite{KCDW}. This simulation assumes $\Omega_o = 0.3$,
$\Lambda=0.7$, $h=0.7$ and $\sigma_8=0.9$. The second most massive cluster
($M_{vir}= 5.6\times 10^{14}h^{-1}M_\odot$) at $z=0$ was placed at the
origin of coordinates and a surrounding spherical region of radius 
$70 h^{-1}$Mpc was isolated for resimulation with Gadget. The 
initial mass distribution between 21 and $70 h^{-1}$Mpc was represented
at relatively low resolution by $3\times 10^6$ particles. In the inner region,
where the original simulation had $2.2\times 10^5$ particles, we created new
initial conditions with $4.5\times 10^5$, $2.0\times 10^6$, $1.3\times
10^7$ and $6.6\times 10^7$ particles. With
increasing mass resolution along this sequence we included extra initial
fluctuations on scales beyond the Nyquist frequency of the
original simulation, and we decreased the force softening to give
improved spatial resolution. We then ran all four simulations to $z=0$
and compared results with each other and with the original simulation.
In the largest resimulation there are about 20 million particles within 
the virial radius of the final cluster and the gravitational softening
radius is $0.7 h^{-1}$kpc. This is the highest resolution simulation
of a rich galaxy cluster ever carried out. 

The particle data for all these simulations were dumped at intervals
of 0.06 in $\ln (1+z)$. We then implemented the semianalytic galaxy
formation recipes of Kauffmann et al \cite{KCDW} on merging trees constructed 
from these outputs. A major improvement is possible as a result of
the increased resolution of the present simulations. The dark halos 
of most of the more massive galaxies remain identifiable as
self-bound substructures within the cluster. 
In the highest resolution simulation there are almost 
5000 such ``galaxy halos'' within the virial radius of the final 
cluster. By keeping track of such substructures we are able to
follow the dynamical evolution of galaxies within clusters and it is
no longer necessary to use semianalytic recipes to track 
galaxy merging. Apart from this change we follow the procedures of
Kauffmann et al exactly, using their ``ejection'' prescription for the
fate of reheated gas, and setting the parameters regulating star 
formation and feedback so that the observed I-band Tully-Fisher
relation and cold gas fraction are reproduced for spiral
galaxies outside the cluster. This requires parameter choices very
similar to those used by Kauffmann et al.

\section{Cluster Results}

We begin by showing that our modelling of galaxy formation, although
adjusted to reproduce the properties of spiral galaxies outside the
cluster, also reproduces well the observed properties of the galaxy 
populations within clusters. The uppermost panel of Fig.~1 compares
the observed Tully-Fisher relation \cite{Gio} with
the predicted relation for isolated Sb/Sc galaxies; the linewidth 
and Hubble type are defined here as in Kauffmann et al \cite{KCDW}. 
The star formation and feedback efficiencies were tuned to
reproduce this relation as well as possible, but it is still striking 
that adjusting two parameters can produce an excellent fit to the slope
and scatter of the observations. It is much more remarkable, however,
that with {\it no} additional parameters the model produces a
cluster luminosity function which is similar in normalisation and 
shape to observed luminosity functions, and a morphology-radius
relation within the cluster which is also like that observed. Note 
that in contrast to earlier work, this latter success is achieved
without tuning any parameter related to merging rates. 

\begin{figure}
\centering
\includegraphics[width=.8\textwidth]{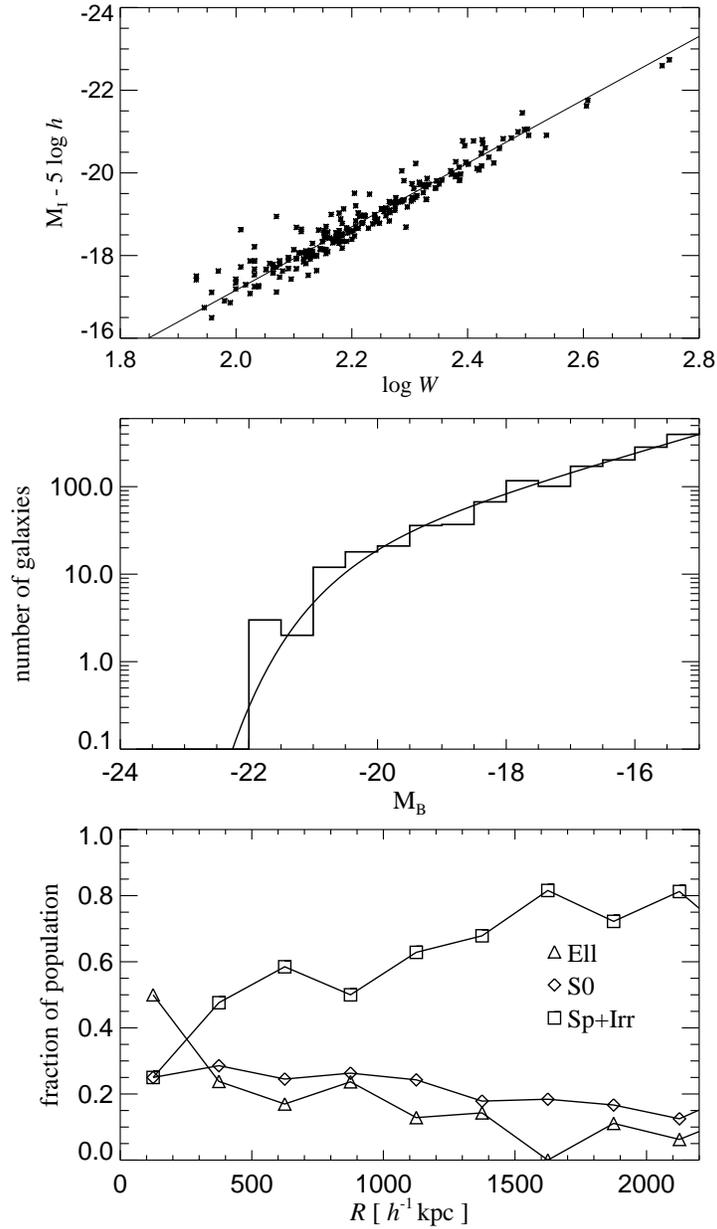}
\caption[]{{\it Top}~~~I-band Tully-Fisher relation for isolated Sb/Sc
galaxies outside the cluster. The straight line is the mean observed relation
from \cite{Gio}. {\it Middle}~~~Number of galaxies within
the virial radius of the cluster as a function of their absolute
magnitude in the B-band. The Schechter fit shown has a faint end
slope, $\alpha=-1.2$. {\it Bottom}~~~Fraction of galaxies brighter
than $M_B=-16.5$ in each of three morphology classes as a function of 
distance from cluster centre.}
\end{figure}

Given that this simulation produces a reasonably good fit to the 
galaxy populations in and around rich clusters, it is interesting to
examine in detail its predictions for the history of star
formation. In our highest resolution simulations the smallest resolved
objects have a total mass of about $5\times 10^8M_\odot$,
corresponding to a total initial baryon mass of $5\times 10^7M_\odot$
although their mass in stars is at least 10 times smaller. The latter 
difference arises because feedback is assumed to make star formation 
inefficient in low mass objects. Thus we resolve the formation of 
stellar systems similar in mass to large globular star clusters.

A first striking effect is the bias towards earlier formation for stars
which end up inside the cluster. Thus half the stars which lie within
the final virial radius form before $z=4$ while half the stars outside
this radius form after $z=2$. The first 1.5\% of stars in cluster 
galaxies have already formed by $z=13$ while one has to wait until 
$z=7.5$ to form a similar fraction of the stars outside the cluster.
Note that formation is still biased in the cluster's vicinity but outside its
virial radius \cite{Lemson}. Thus the first 1.5\% of the
stars in a ``typical'' region of the Universe would form even later.

In the upper right panel of Fig.~2 we show a projection at $z=13$ 
of the galaxies in a cube of comoving side $10h^{-1}$Mpc centred on 
the barycentre of the material of the future cluster. The area of 
each symbol is proportional to the stellar mass of the galaxy. This
distribution is clearly extremely inhomogeneous, with almost all the
stars lying on or close to a single filament of comoving scale about 
$10h^{-1}$Mpc. Clearly the heavy elements produced by these ``first stars''
are very unlikely to be well mixed through the IGM. In the lower left
panel we show where these same stars are at $z=0$; the area of
each symbol is here proportional to the mass of ``first stars'' which the
corresponding galaxy contains. This can be compared to the plot at
upper left which shows the distribution of all the stars at the end of
the simulation, symbol area now corresponding to total stellar mass.
Clearly the first stars are concentrated in a relatively small number
of galaxies and these galaxies tend to lie close to the cluster
centre. In fact, at $z=0$ more than half of all the ``first stars'' lie
within $100h^{-1}$kpc of cluster centre, whereas fewer than 10\% of
all cluster stars lie at such small radii.

Because of these strong bias effects the first stars in different
regions of the Universe are predicted to form at very different times;
some may still be forming today in weak structures in ``voids''. For
many purposes it is more interesting to define the first stars not
as those which formed first in time, but as those which formed from
the least processed gas. In hierarchical cosmogonies the degree of
processing of gas is closely related to the mass of the object in which
it is found. For example, in the simulations analysed here gas which
has never been part of a virialised object at all, or never part of an
object with virial temperature exceeding $\sim 10^4$K, has never been
associated with star formation and could plausibly have zero heavy
element abundance. Such effects may be reflected in the strong
observed correlation between the mean metal abundance of galaxies and their
stellar mass. We illustrate the possible distribution of the ``lowest
metallicity'' stars in the lower right panel of Fig.~2. This shows
the distribution at $z=0$ of the 1.5\% of cluster stars which formed
in the lowest mass dark halos. These stars are still primarily part of
the lowest mass galaxies and their radial distribution within the
cluster is similar to that of all the stars. Half of this subset of
stars formed after $z=3.9$, and their age distribution is
very similar to that of clusters stars as a whole. Low
metallicity stars in clusters are no older than the average. 

\begin{figure}
\centering
\includegraphics[width=\textwidth]{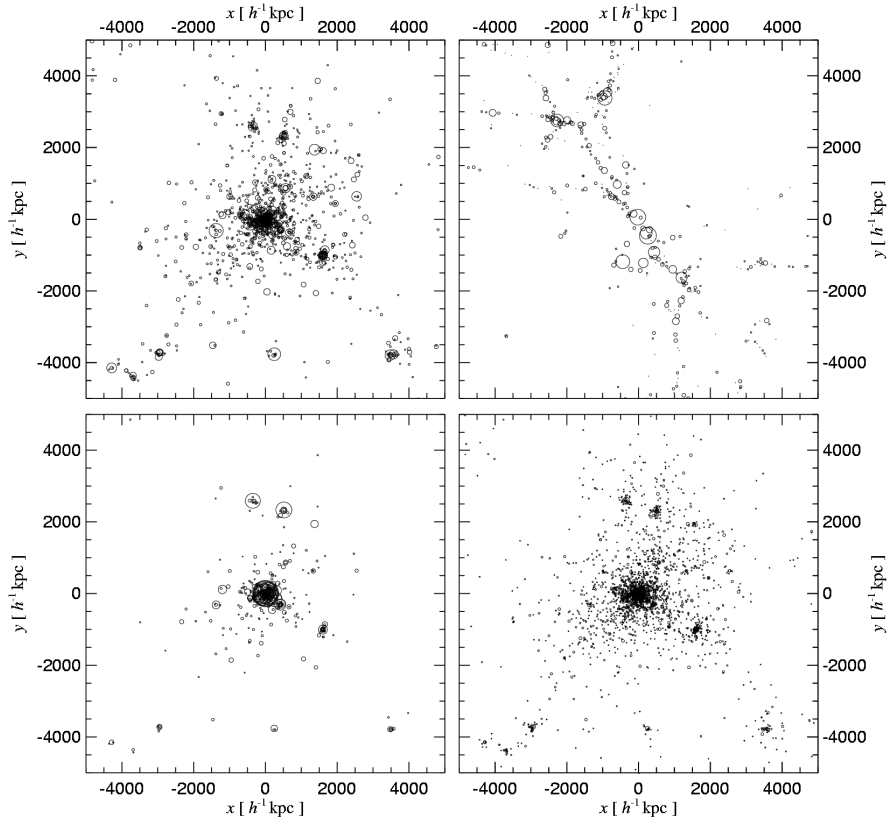}
\caption[]{Projections of the galaxy distribution in a cube of
comoving side $10h^{-1}$Mpc. The upper right image is at $z=13$ when
1.5\% of the stars in final cluster have formed. The other three
images are at $z=0$. The upper images show all galaxies in the cube
with symbol area proportional to stellar mass. The lower two images
show only galaxies containing stars which were among either the first 
1.5\% (left) or the 1.5\% formed in the lowest mass halos (right). In these
lower images symbol size is proportional to the mass of the relevant
stellar population.}
\end{figure}

\section{``Milky Way'' Results}

Rather than running a new simulation of the assembly of a ``Milky
Way'' halo, we here take a short cut by scaling our cluster
simulations down by a factor of 7 in both size and velocity,
corresponding to a factor of 343 in mass and to unchanged dynamical
times. Both theoretical \cite{LC} and numerical \cite{Moore}
results suggest that such a scaling should give a good 
approximation to the dark matter evolution of a ``typical'' halo, 
except that assembly is shifted to somewhat lower redshifts. We run 
our semianalytic machinery on these rescaled simulations (using the 
same parameters as before) to obtain predictions for the star
formation history within a ``Milky Way'' halo.

The results are illustrated in Fig.~3 which can be compared directly
with Fig.~2. The shift in temperature (by a factor 49) caused by
our rescaling produces large changes in cooling and feedback.
Most star formation occurs at late times in the central disk. 
Within the final galaxy's halo about 80\% of all stars are in the 
disk, 16\% are in the bulge, and only 3\% are part of other objects, 
more than half of those in a single satellite. 
Only by $z=6.9$ has our model Milky Way made
1\% of its stars. Figure 5 of ref. \cite{NFW} suggests that this
should be increased to $z\sim 10$ to account for the bias towards
late assembly just noted. Despite the much higher effective resolution
of the rescaled simulation, this redshift is still well below the 1\%
formation redshift of stars in the cluster of the last section.

The upper right panel of Fig.~3 shows the initial distribution of
these ``first'' stars to be very inhomogeneous -- most are in a handful of
progenitor galaxies. By $z=0$ they are quite centrally concentrated;
60\% are within 10 kpc, with even older stars being even more centrally
concentrated. On the other hand, if low metallicity stars are 
identified as those formed in the lowest mass dark halos, only 16\% of
the lowest metallicity 1\% lie within 10 kpc, and lower metallicity 
stars are predicted to be even less centrally concentrated.
These ``low metallicity'' stars have a median formation redshift 
of $z=5.5$, much higher than the formation redshift of ``typical'' stars.
Most are part of intact (60\%) or
disrupted (30\%) dwarf satellites; only 10\% are part of the central
bulge. We return to these numbers in the next section.

\begin{figure}
\centering
\includegraphics[width=\textwidth]{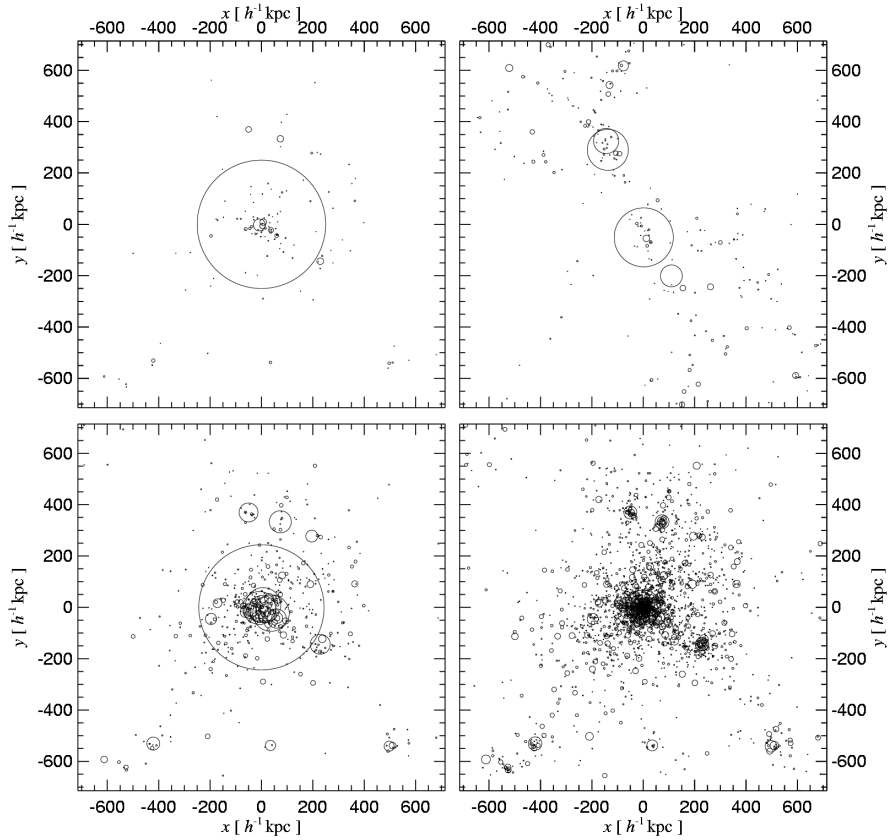}
\caption[]{Projections of the galaxy distribution in a cube of
comoving side $1.4h^{-1}$ Mpc around a ``Milky Way'' halo. The upper right
image is at $z=5.4$ when 1\% of the final stars have
formed. The other three are at $z=0$. The illustrated quantities are
as in Fig.~2 except that the lower plots show the 1\%
``earliest'' and 1\% ``lowest metallicity'' stars.}
\end{figure}

\section{Discussion}

The results shown above demonstrate that if current structure
formation theories are correct, the oldest stars should be found
at the centres of rich clusters. Furthermore the distribution of the
first stars was inhomogeneous on the (comoving) scale of {\it
present-day} large-scale structure. As a result, enrichment of
the intergalactic medium at high redshift must have been extremely
patchy. If we want to find the oldest stars in the local universe we
should look near the centre of M87, not in our own Galactic halo or in
intergalactic space. Unfortunately they may be indistinguishable
from the enormous background of slightly younger stars. On the 
other hand, the lowest metallicity stars (assumed here to be those 
that formed in the lowest mass objects) are distributed on large
scales in a similar way to the general stellar population.

On smaller scale, similar effects imply that the classic monolithic
collapse \cite{ELS} and inhomogeneous assembly \cite{SZ} pictures are
of little help in interpreting the age and metallicity distribution
within the Galactic bulge and halo. The oldest stars
are predicted to be in the inner halo or bulge, and will be very difficult to
distinguish from the dominant, somewhat younger population. 
Lower metallicity populations are expected to have more extended 
distributions, but there is no strong correlation between metallicity 
and age. Some low metallicity stars may be quite young, and
indeed may continue to form today in low mass, isolated dwarf galaxies. 

The rich internal structure predicted for dark matter halos in these
CDM models is, as Fig.~1 demonstrates explicitly, in very satisfying
agreement with the observed galaxy populations in rich clusters.
Ben Moore and his collaborators \cite{Moore} have emphasised that the
situation is less clear for the Milky Way. Most 
of the low metallicity stars in our simulation are in the stellar halo
rather than the bulge, 
but twice as many reside in satellites as in the diffuse component.
In the real Galaxy more are in the diffuse halo than in satellites. 
This may be because the cores of
many low mass halos survive undisrupted in our model, while the Milky 
Way can disrupt dwarfs even from relatively
weakly bound orbits, as the 
example of Sagittarius clearly shows \cite{Ibata}. It is uncertain
whether this is a serious shortcoming of CDM models or may
be resolved by more detailed dynamical modelling of the formation
and evolution of satellites. The luminosity function we predict for
satellites is {\it not} in obvious conflict with observation.

In our Milky Way model most of the spheroidally distributed
stellar mass is contributed by the bulge of the central galaxy. As in
other hierarchical models, these 
stars formed in the gaseous disks of a small number of relatively
massive progenitors prior to or during their merger, and before
the formation of the present disk \cite{Kauffmann}. The 
disruption of Sagittarius-like dwarfs contributes relatively few
stars to the bulge.

If the stellar spheroid of the Milky Way was indeed assembled in this
way, one may ask whether its current structure
should retain any trace of its inhomogeneous origin. Inside the Solar
circle, where most of the spheroid stars reside, orbital times are so
short that the spatial distribution of debris from a 10 Gyr old
merger is expected to show few sharp features. In phase-space,
however, it may still occupy a small fraction of the accessible
region, being confined to a kinematically cold sheet which wraps many
times around the Galaxy. Such structure may be detectable with
sufficiently good kinematic data; the stars in the neighborhood of
any given point will not be distributed uniformly through velocity
space but will be confined to a few hundred cold streams \cite{Helmi}.
A recent analysis of the velocities of nearby metal-poor giants, based
on combined ground-based and Hipparcos data, has detected two such 
streams in the Solar neighborhood \cite{Helmi1}. These appear to be
debris from a galaxy like Fornax or Sagittarius, destroyed
roughly 10 Gyr ago. This is, perhaps, the first direct
evidence for hierarchical assembly of the stellar halo.

\vspace{0.1cm}
We thank G.~Tormen for help in creating initial conditions 
and G.~Kauffmann for help in implementing semianalytic
recipes.


\clearpage
\addcontentsline{toc}{section}{Index}
\flushbottom
\printindex

\end{document}